\documentclass[10pt]{iopart}

\usepackage{iopams}  
\usepackage{graphicx}
\usepackage{verbatim}
\usepackage{xcolor}
\usepackage{subfig}
\usepackage[export]{adjustbox}
\begin{document}

\title[Multiple dynamic regimes in a coarsening foam]{Multiple dynamic regimes in a coarsening foam}

\author{Fabio Giavazzi$^{1}$, Veronique Trappe$^{2}$, and Roberto Cerbino$^{1}$}

\address{1 Dipartimento di Biotecnologie Mediche e Medicina Traslazionale,
Universit\`a degli Studi di Milano, via F.lli Cervi 93, 20090 Segrate,
Italy}
\address{2 Department of Physics, University of Fribourg, Chemin du Musée 3,
CH-1700, Fribourg, Switzerland}

\eads{\mailto{fabio.giavazzi@unimi.it},\mailto{veronique.trappe@unifr.ch},\mailto{roberto.cerbino@unimi.it}}\vspace{10pt}
\begin{indented}
\item[]June 2020
\end{indented}

\begin{abstract}
We use differential dynamic microscopy and particle tracking to determine the dynamical characteristics of a coarsening foam in reciprocal and direct space. At all wavevectors $q$ investigated,  the intermediate scattering function exhibits a compressed exponential decay. However, the access to unprecedentedly small $q$s highlights the existence of two distinct regimes for the $q$-dependence of the foam relaxation rate $\Gamma (q)$.  At any given foam age, $\Gamma (q)\sim q$ at high $q$, consistent with directionally-persistent and intermittent bubble displacements. At low $q$, we find $\Gamma (q) \sim q^{1.6}$. We show that such change in $q$-dependence of $\Gamma (q)$ relates to a bubble displacement distribution exhibiting a cut-off length of the order of the bubble diameter. Investigations of the $q$-dependence of $\Gamma (q)$ at different foam ages reveal that foam dynamics is not only governed by the bubble length scale, but also by the strain rate imposed by the bubble growth; normalizing $\Gamma (q)$ by this strain rate and multiplying $q$ with the age-dependent bubble radius leads to a collapse of all data sets onto a unique master-curve.\end{abstract}

%
%
\submitto{\JPCM}
%
\maketitle
%
\ioptwocol
\section{Introduction}
Liquid foams commonly consist of polydisperse gas bubbles that are highly packed in a liquid continuous phase \cite{weaire2001physics}. As the bubble packing fraction generally exceeds random close packing, the bubbles exert direct contact forces onto one another. To maintain a static bubble configuration, these forces need to be balanced.  However, because of differences in Laplace pressures between bubbles of different size, foams coarsen in time, the large bubbles growing at the expense of small ones.  This coarsening process continuously alters the stress configuration of the system, leading to locally imbalanced stresses that in turn trigger local bubble rearrangement events \cite{Durian1990coarsening,durian1991multiple}.   

Such stress-driven dynamics has been inferred to be at the origin of residual activity in a number of aging soft matter systems \cite{ramos2001ultraslow,cipelletti2003universal}. Experimental evidence for this scenario were obtained in dynamic light scattering experiments, yielding intermediate scattering functions displaying compressed exponential decays with relaxation rates $\Gamma (q)$ depending linearly on the scattering wavevector $q$, reminiscent of ballistic motion \cite{cipelletti2003universal,cipelletti2005slow,li2019physical}. To account for this behavior it was proposed that randomly distributed dipolar stress sources generate displacement fields that lead to directionally-persistent displacements characterized by a power-law tailed probability distribution \cite{bouchaud2001anomalous,duri2006length,bouchaud2008anomalous}, and numerical studies suggested that this could be the case in a broad range of systems exhibiting stress relaxation \cite{ferrero2014relaxation,bouzid2017elastically,pelusi2019avalanche}. Though this conjecture is appealing, direct experimental evidence of the link between stress-induced displacements and reciprocal space characteristics is to date very limited \cite{tamborini2014plasticity}.

In this work, we aim to fully explore this link in foams. This choice is motivated by the 
fact that the source of stress imbalances in foams can be traced back to the continuous bubble growth, and because the size of the foam bubbles is sufficiently large to use microscopy as a main investigative tool. Microscopy image sequences of the foam acquired during coarsening are analyzed both with particle tracking and with differential dynamic microscopy \cite{cerbino2008differential,giavazzi2009scattering,giavazzi2014digital,cerbino2017perspective} to obtain a comprehensive set of data probing foam dynamics in both direct and reciprocal space. 

Our experiments reveal that foam dynamics is governed by intermittent bubble displacements exhibiting a persistent direction up to the bubble length scale. This length scale introduces a cut-off in the probability distribution function of the bubble displacements that otherwise exhibits the power-law scaling expected  \cite{bouchaud2001anomalous,duri2006length,bouchaud2008anomalous}. We demonstrate experimentally and theoretically that such cut-off leads to distinct $q$-dependencies of the relaxation rates depending on whether $1/q$ larger or smaller than the cut-off length. 

Moreover, we find that the dispersion relations obtained at different foam ages collapse onto a unique master curve, by rescaling $q$ with the bubble size and $\Gamma (q)$ with the coarsening strain rate $\Gamma_{c}=\dot{R}/R$. This shows that foam dynamics is uniquely ruled by a single length and time scale, imposed by respectively the foam structure and coarsening kinetics.

\section{Materials and methods}
\subsection{Sample preparation and imaging \label{sec:sample}}
Our sample is a commercial shaving foam (Gillette Foamy Regular), which has been previously shown to exhibit reproducible coarsening characteristics, and to be reasonably stable against coalescence and drainage \cite{Durian1990coarsening}.   

The start of our experimental time frame is set by the foam production ($t_{w}$ = 0), at which point the foam is directly injected into a two-piece polystyrene Petri dish (radius 35 mm and height 10 mm). The Petri dish is subsequently sealed with Parafilm and immediately transferred to the microscope. 

Images of the bubble layer in contact with the bottom side of the Petri dish are taken in back reflection, using a $2\times$ objective with a numerical aperture of $NA = 0.06$. The microscope used is an inverted microscope (Nikon Eclipse Ti-E) equipped with a digital camera (Hamamatsu Orca Flash 4.0 v2), epi-illumination being provided by a blue LED (Thorlabs M455L4-C, peak wavelength 455 nm). The pixel size is 13 $\mu$m. The images are 3.3 x 3.3 mm$^2$ in area and contain 4000 - 400 bubbles depending on the age of the foam. We acquire images over a period of $14$ hours at a frame rate $1/\Delta t_{0}$ of $1$ fps.

\begin{figure*}[!th]
\centerline{\includegraphics[width=1.15\linewidth]{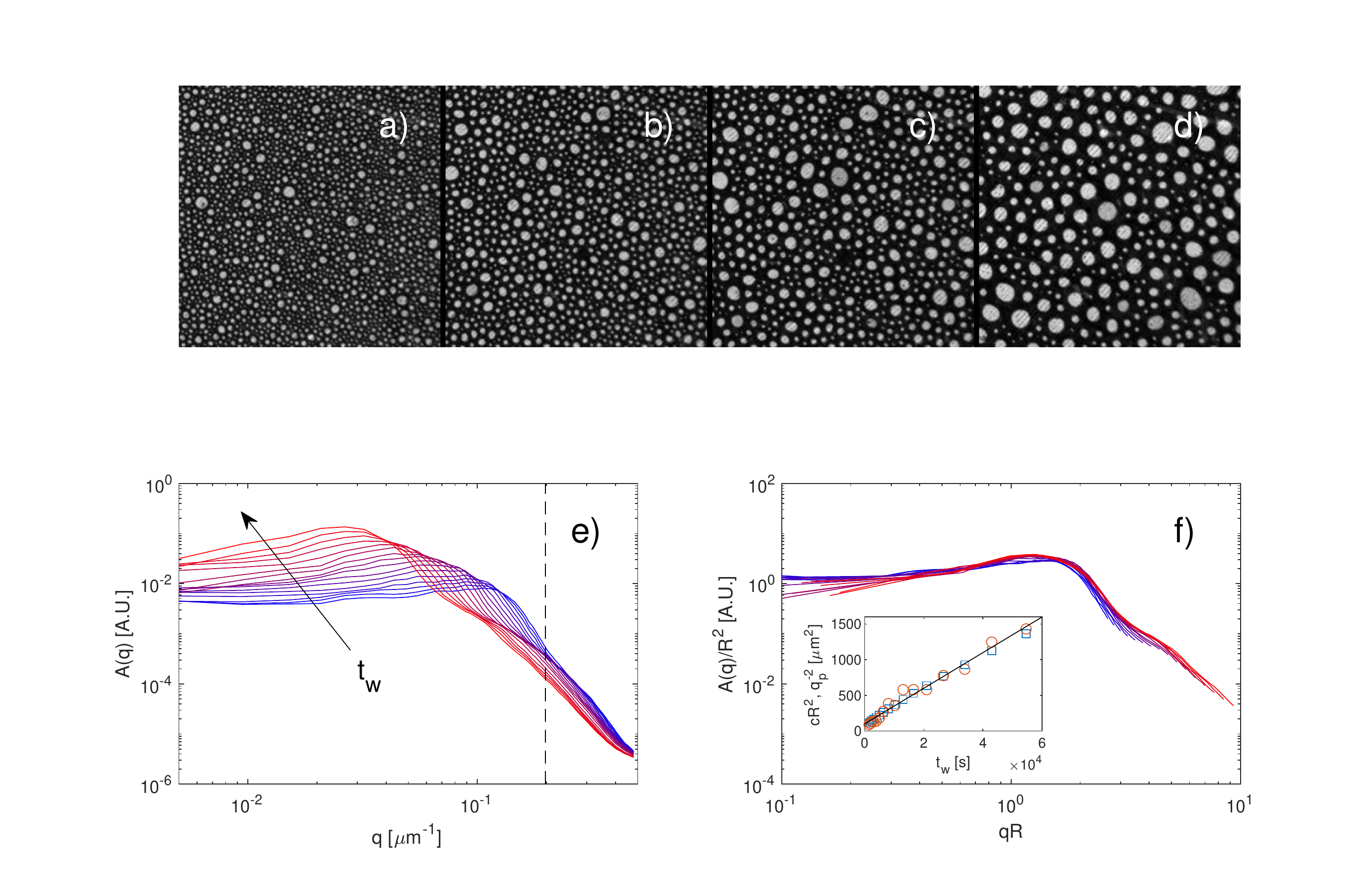}}
\caption{\label{fig1}
Structural characteristics of coarsening foam. a-d) Images  taken at {$t_w=1000, 3000, 5000, 8000, 20000$} s. The size of each image corresponds to 2 x 2 mm$^2$ in real space. e) Azimuthally averaged Fourier power spectra $A(q)$ obtained at $t_w$ ranging from 1600s to 54000s. The vertical dashed line denotes the limit beyond which the $A(q)$ is distorted by artifacts due to the microscope transfer function f) Scaled representation of the data shown in panel e); the $q$ axis is scaled with the mean bubble radius $R(t_w)$, $A(q)$ is scaled with $R(t_w)^{-2}$. Inset: orange circles denote the age dependence of the inverse squared peak position of $A(q)$, $q_{p}(t_w)^{-2}$; blue squares denote the age dependence of the squared bubble radius $R^2$ multiplied with a constant $c=0.60$. The continuous line corresponds to the best linear fit to the data.}
\end{figure*}

Representative cropped images taken at different times after production $t_{w}$ are shown in Fig.\ref{fig1} a-d, where the gas bubbles appear bright and the continuous phase dark. Optical contrast is here mainly generated by the reflectivity of the interface between the gas bubbles and the continuous phase. 

To determine the age-dependent foam dynamics at quasi-stationary conditions, the entire image sequence is divided into many partially overlapping sub-sequences that are analysed separately. The $n$-th sub-sequence $S_n$ is centered at age $t_w=10^3\exp{[n/4]}$ s $(n=1, 2,..., 16)$ and covers a time interval of $t_w/4$. This choice warrants that the mean bubble size changes by less than 15\% within each sub-sequence.

\subsection{\label{sec:rec_space} Reciprocal space analysis}
The foam structure and dynamics are characterized in reciprocal space by using the differential dynamic microscopy (DDM) protocol   \cite{cerbino2008differential,giavazzi2009scattering,giavazzi2014digital}. For each sequence $S_n$, we determine the azimuthally averaged Fourier power spectrum (static amplitude) $A_{n}(q)$, and the intermediate scattering function $f_{n}(q,\Delta t)$. To simplify the notation, the index $n$ referring to the image sequence is omitted in the following. 

In a preliminary step, we remove the effect of uneven illumination by dividing each image by the background image $I_b(\mathbf{x})$ obtained by applying a Gaussian filter with standard deviation $200$ $\mu m$ to the temporal average of all the images in the sequence.

We then calculate the difference between two background corrected images acquired at times $t$ and $t+\Delta t$, $d(\mathbf{x},t,\Delta t)=I(\mathbf{x},t+\Delta t)-I(\mathbf{x},t)$. By averaging the spatial Fourier power spectrum of $d(\mathbf{x},t,\Delta t)$ obtained for the same $\Delta t$ but different reference times $t$ we obtain the \textit{image structure
function} $d(\mathbf{q},\Delta t)=\langle |\hat{d}(\mathbf{x},t,\Delta t)|^{2}\rangle _{t}$ that captures the sample dynamics as a function of the two-dimensional scattering wavevector $\mathbf{q}$ and of the lag time $\Delta t$. The symbol $\hat{\cdot}$ indicates the two dimensional digital Fourier transform, usually performed with a Fast Fourier Transform algorithm. The average taken over different $t$ is justified because the dynamics and structure are quasi-stationary within the time interval at which an image sub-sequence is taken. 

In a last step, we take advantage of the circular symmetry of the image structure function to perform an azimuthal average of $d(\mathbf{q},\Delta t)$, which provides the dimensionally-reduced structure function $d(q,\Delta t)$ of the radial wavevector
$q=\sqrt{q_{x}^{2}+q_{y}^{2}}$. 

This structure function is connected to the intermediate scattering function (ISF) $f(q,\Delta t)$ \cite{berne2000dynamic} by the relation \cite{giavazzi2014digital}
\begin{equation}
d(q,\Delta t)=2A(q)\left[1-f(q,\Delta t)\right]+2B(q)\label{eq:strufu}
\end{equation}
with $B(q)$ accounting for the camera noise and the \textit{static amplitude} $A(q)=T(q)I(q)$, depending on the static scattering intensity of the sample $I(q)$ and the transfer function of the microscope $T(q)$.

For the images taken in our experiment, the main contribution to random fluctuations comes from shot noise, which is delta-correlated in space. This entails that the noise term $B(q)$ is practically $q$-independent.  Since $A(q \rightarrow \infty)=0$ and $f(q,\Delta t \rightarrow 0)=1$ due to the finite numerical aperture of the objective, we can estimate the magnitude of $B$ as the high-$q$, small $\Delta t$ limit of $d(q,\Delta t)$ \cite{giavazzi2009scattering}.

Once $B$ is known, we could in principle use Eq. \ref{eq:strufu} to extract both $A(q)$ and $f(q,\Delta t)$. However, because of the limited time interval over which an image sub-sequence is taken, a full relaxation of $f(q,\Delta t)$ is not observed at all $q$ values. To overcome this limitation, we estimate $A(q)$ from the time-averaged power spectrum of the individual images as $A(q)\simeq\langle|\hat{I}(\mathbf{q},t_0)|^2\rangle_{|\mathbf{q}|=q,t_0}-B(q)$ \cite{cerbino2017dark}. For our experiment, this approximation is justified as the optical signal produced by the foam is much stronger than any contribution of stray light or dirt on the optical components.
\subsection{\label{sec:dir_space}Direct space analysis}
To characterize the bubble dynamics in direct space, we apply a particle tracking (PT) analysis on the image sub-sequence corresponding to the age $t_w=1.5 \cdot 10^4$ s. As done for the DDM analysis, we first correct the images for uneven illumination. The background-corrected images 
are then filtered with a Gaussian kernel with standard deviation $5$ $\mu m$ to reduce noise, and subsequently converted into binary masks by applying a fixed threshold value.     

Bubbles are identified as connected regions $B_m(t)$ of the binary mask with surface area $a_m(t)$ larger than a fixed cutoff value of 120 $\mu m^2$. The bubble center $\mathbf{x}_m(t)$ is determined as the intensity-weighted center of mass of the corresponding region in the compensated image $\mathbf{x}_m(t)=\sum_{x \in B_m(t)}{ \left[ \mathbf{x} \cdot I_c(\mathbf{x},t) \right] } /\sum_{x \in B_m(t)}{I_c(\mathbf{x},t)}$.

The typical displacement of a bubble between consecutive frames is well below the pixel size, while the largest displacement observed is always smaller than the bubble diameter. We can thus link the position of each bubble in two consecutive frames by maximizing the overlap between the respective surface areas to reliably determine the single bubble trajectories. 

Once the trajectories of all bubbles are available, we evaluate the probability distribution function (PDF) $P(\Delta r|\Delta t)=\langle \delta \left(\Delta r - |\mathbf{x}_m(t+\Delta t)-\mathbf{x}_m(t)| \right) \rangle$ of the bubble displacement, where the average is calculated over all bubbles $m$ and initial times $t$. In practice, $P$ is evaluated for each $\Delta t$ as the normalized frequency histogram of $|\mathbf{x}_m(t+\Delta t)-\mathbf{x}_m(t)|$ with logarithmic binning using 40 bins covering the interval 0.01 -10 $\mu m$.

The mean square bubble displacement (MSD) is determined as $r(\Delta t)=\langle |\mathbf{x}_m(t+\Delta t)-\mathbf{x}_m(t)|^2 \rangle$,
where the average is again performed over all bubbles $m$ and initial times $t$.

The characteristic foam bubble radius $R(t_{w})$ at a given age is estimated from the relation $R^2(t_{w})=\langle a_{m}(t_w) \rangle/\pi$.

\section{Results and discussion}
\begin{figure*}[!th]
\centering
\subfloat{\includegraphics[clip,width=0.95\linewidth,valign=t]{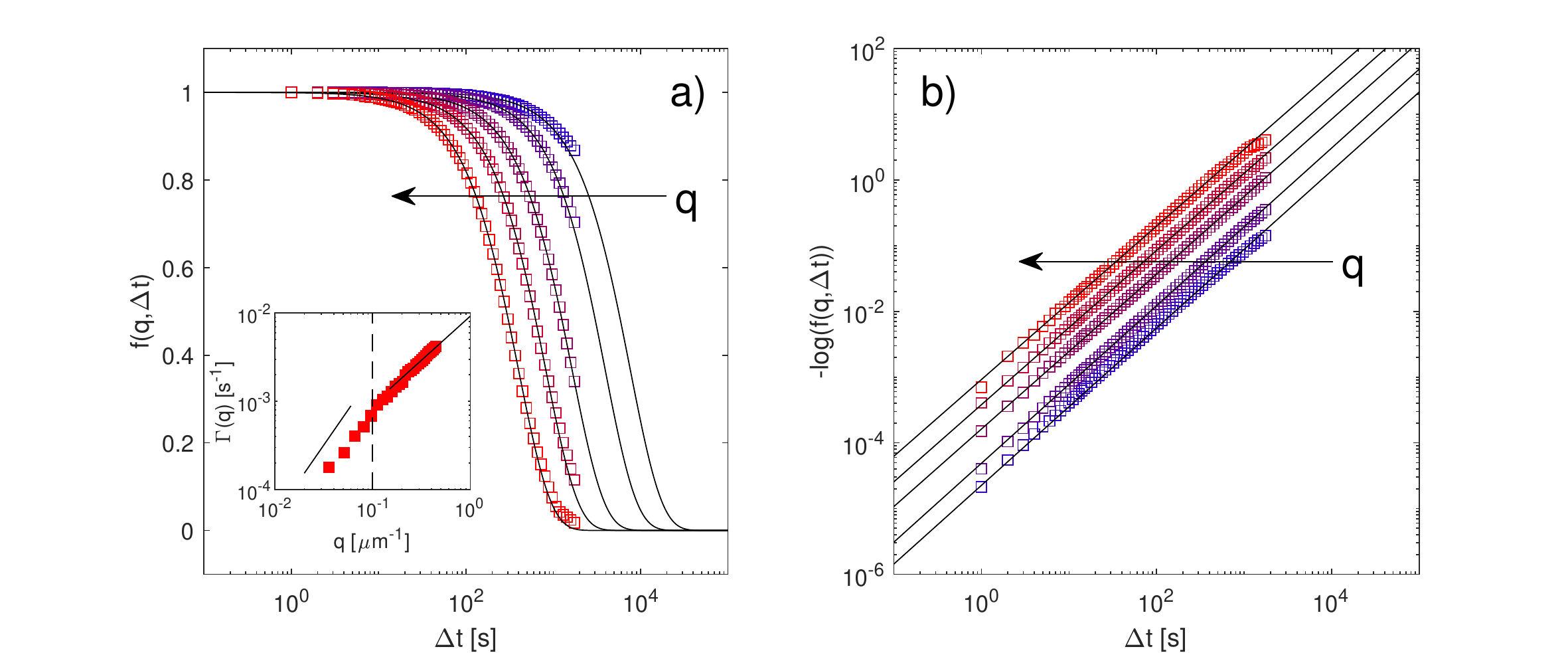}}
\newline
\subfloat{\includegraphics[clip,width=0.95\linewidth,valign=t]{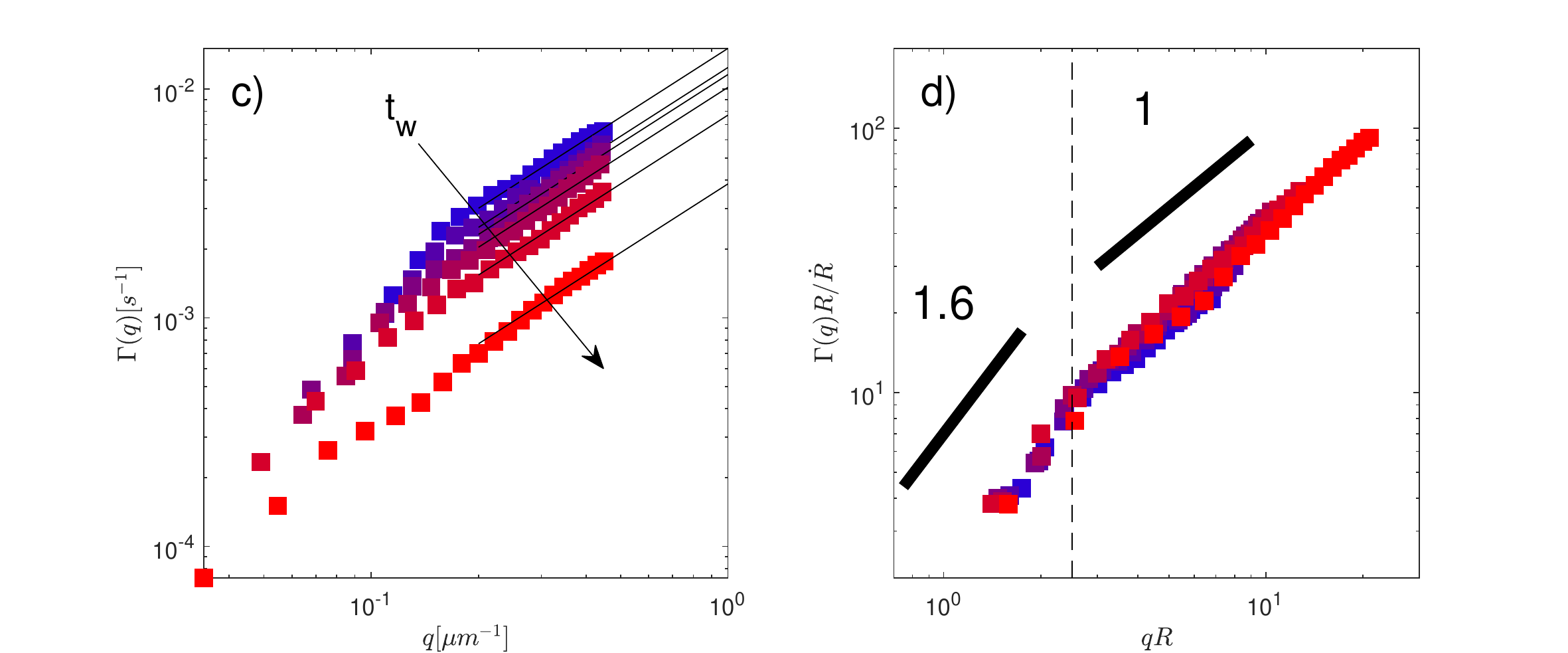}}
\caption{\label{fig2}
Reciprocal space analysis of foam dynamics. a) Intermediate scattering functions $f(q,\Delta t)$ obtained at a foam age of $t_w=10000$ s for $q$-values covering the range of 0.02-0.3$\mu m^{-1}$. Continuous lines are fits to the data using compressed exponentials of form $\exp{\{-(\Gamma(q)\Delta t)^{\alpha}\}}$, with $\alpha \simeq 1.2$. Inset: $q$-dependence of the relaxation rates $\Gamma(q)$ obtained from the fits. b) Logarithm of the ISFs shown in panel a). Continuous lines are best fits to the data using power-laws of form $(\Gamma(q)\Delta t)^{\alpha}$. c) $q$-dependence of relaxation rates obtained at different foam ages. Continuous lines are best fits to the large $q$-limits using a linear model of form $\Gamma(q)=u_0 q$.  d) Scaled representation of the data shown in panel c). A master curve is obtained by scaling $q$ with $R(t_w)$ and normalizing $\Gamma(q)$ with the coarsening rate $\dot{R}(t_w)/R(t_w)$. The vertical dashed line corresponds to the rescaled cross-over wave-vector $q_{c}R\simeq 2.4$ separating the low-$q$ regime, where $\Gamma(q)\sim q^{1.6}$, from the "ballistic" regime at larger $q$, where $\Gamma(q)\sim q$.}
\end{figure*}

\subsection{\label{sec:statics} Reciprocal space: foam structure}
Because of the differences in Laplace pressure between small and large bubbles, the average bubble size of our foam increases with increasing $t_{w}$ (Figure \ref{fig1}a-d). Such evolution reflects in a change of the $q$-dependence of the static amplitude $A(q)=T(q)I(q)$. As shown in Figure \ref{fig1}e, $A(q)$ is characterized by a well-defined peak, which shifts towards lower $q$-values with increasing $t_{w}$. In our experiment, $T(q)$ is almost constant up to $q\simeq 0.2 \mu m^{-1}$, such that the $q$-dependence of $A(q)$ essentially reflects that of $I(q)$ below that $q$-value. Considering only the low $q$-range, a simple normalization of $A(q)$ by $R(t_{w})^2$ and $q$ with $R(t_{w})$ leads to a good collapse of all data-sets onto a unique master-curve, as shown in Figure \ref{fig1}f. This denotes that the mean bubble radius is the only parameter that varies during coarsening, the average bubble configuration remaining essentially the same.

For dry foams we expect a linear growth of the bubble area with time \cite{weaire2001physics}. This is consistent with our experimental results. As shown in the inset of Figure \ref{fig1}f, the square of the bubble radius or equivalently the square of the inverse scattering vector at the peak of $A(q)$, $q_{p}$, depend linearly on $t_w$. Fitting the data for $q^{-2}_{p}(t_w)$ with a linear function of form $q^{-2}_{p}(t_w)=q^{-2}_{p}(0)+Kt_w$ yields $K=(2.5 \pm 0.05) \cdot 10^{-2}$ $\mu$m$^2/s$ for the average coarsening constant. A fit to a more general model, $q^{-2}_{p}(t_w)=q^{-2}_{p}(0)+K_\beta t^{\beta}_w$, provides a slightly different scaling exponent $\beta=0.84 \pm 0.05$. This deviation from an "ideal" coarsening behavior is likely due to a small drainage-induced increase of the liquid fraction at the bottom of the cell, which is the plane we observe experimentally.

\subsection{\label{sec:dynamics} Reciprocal space: bubble dynamics}
To assess the impact of coarsening on the rearrangement dynamics of the foam, we analyze the intermediate scattering function $f(q,\Delta t)$ for different foam ages. As a representative example, we show the ICFs obtained for $t_w=10000$s in Figure \ref{fig2}a. As we restrict the determination of $f(q,\Delta t)$ to a time window over which we can expect the dynamics to be quasi-stationary, the accessible data range is limited and a full decay of $f(q,\Delta t)$ is only obtained for large $q$-values. However, despite this limitation we can assess the decay rate of $f(q,\Delta t)$ by fitting the initial decay as $f(q,\Delta t) = \exp{[-(\Gamma(q)\Delta t)^{\alpha}]}$.  This can be appreciated in Figure \ref{fig2}b, where we report the dependence of $ -\ln {f(q,\Delta t)} $ as a function of $\Delta t$ in a double logarithmic plot. At all $q$-values investigated, the initial slope is described by a unique value corresponding to $\alpha \simeq 1.2$, while the absolute values of $ -\ln {f(q,\Delta t)} $ shift towards lower $\Delta t$ as $q$ increases, reflecting the increase of the relaxation rate .      
Remarkably, we find that the $q$-dependence of the relaxation rate  shows two distinct regimes. As shown in the inset of Figure \ref{fig2}a, $\Gamma(q)$ scales linearly with $q$ in the range of $q \gg 0.1$ $\mu m^{-1}$, while for low $q$ a stronger dependence is found, $\Gamma(q) \sim  q^{\delta}$ with $\delta \approx 1.6$.

The analysis at different foam ages yield similar results: the ISFs are well described by compressed exponential functions with a compressing exponent $\alpha \simeq 1.2$; the $q$-dependent relaxation rates displays two dynamic regimes separated by a crossover scattering vector $q_c$. However, $q_c$ progressively shifts towards lower $q$-values as the foam coarsens (Figure \ref{fig2}c). In addition, we find that the prefactor $u_0$ of the linear scaling regime $\Gamma(q)\simeq u_0 q$, becomes markedly smaller as the foam ages.

Remarkably, a simple normalization of the horizontal axis of the dispersion relation $\Gamma(q)$ with the characteristic bubble radius $R(t_w)$ and the vertical axis with the strain rate $\dot{R}(t_w)/R(t_w)$ associated to coarsening leads to a collapse of all data sets onto a single master curve (Figure \ref{fig2}d). 

This scaling denotes that foam dynamics is determined by a single length and a single time scale, the bubble size and the strain rate associated to the coarsening process, respectively.

\subsection{\label{sec:model} A simple model accounting for the dynamical characteristics in reciprocal space}

At large enough $q$, our results denote that compressed exponential relaxations are associated to a ballistic-like dispersion relation. Such combination has been found in a variety of non-equilibrium systems \cite{cipelletti2003universal,cipelletti2005slow,ruta2012atomic,angelini2013dichotomic,orsi2014controlling,ferrero2014relaxation,gao2015microdynamics,li2019physical}, and has been rationalized in terms of a heuristic model, originally developed for colloidal gels \cite{cipelletti2000universal} \cite{bouchaud2001anomalous}. This model entailed that randomly distributed local rearrangement events would lead to stress inhomogeneities that act as dipolar forces, inducing strain fields that would give rise to ultraslow yet continuous ballistic-like motion of the gel strands. Additional work then indicated that intermittent rearrangement events could be at the origin of a linear dispersion relation provided that they would lead to displacements with directional persistence \cite{duri2006length}.  

Independent of the actual physical origin, the general idea is that a compressed exponential decay of the self intermediate scattering function $f_{s}(q,\Delta t)=e^{-(\Gamma(q) \Delta t)^\alpha}$, with $\Gamma(q)=u_0 q$, results from a probability density function of particle displacements $P(\Delta r, \Delta t)$ that exhibits a power-law tail $\sim  \Delta r^{-(\alpha+1)}$. 
This can be understood as follows. The Fourier transform of a compressed exponential function of form $g(u)=e^{-|u|^\alpha}$ is the Levy stable distribution $L_{\alpha,0}(u)$, which displays a power-law tail for large values of its argument $L_{\alpha,0}(u)\sim |u|^{-(\alpha+1)}$ \cite{cipelletti2003universal}. 
For a one-dimensional system with a self-ISF of form $f_s(q,\Delta t)=e^{-(u_0\Delta t q)^{\alpha}}$, this entails that the spatial Fourier transform of $f_s(q,\Delta t)$, which corresponds to the PDF of the particle displacements,
will be given by $P(\Delta r, \Delta t)=\frac{1}{u_0 \Delta t} L_{\alpha,0}(\frac{\Delta r}{r_0 \Delta t })\sim (u_0 \Delta t)^{\alpha} \Delta r^{-(\alpha+1)}$.

As demonstrated explicitly for the 3D case in Ref. \cite{cipelletti2003universal}, this result can be generalized to an arbitrary space dimension $d$, showing that compressed exponential relaxations of ISFs always imply the presence of power-law tails in the PDF of particle displacements.
However, let us note that when the compressing exponent $\alpha$ is smaller than 2, the particle mean square displacement $\langle \Delta r^2 \rangle=\int^{+\infty}_0 {r^2P(r, \Delta t)dr}$ is infinite for every $\Delta t$. This rather unphysical situation can be mitigated by assuming the existence of some physical cut-off length $l_0$ limiting the maximum displacement of the particles. If this is the case, the MSD becomes finite $\langle \Delta r^2 \rangle \sim l^2_0(u_0(\Delta t)/l_0)^{\alpha}$.

The introduction of a cut-off length $l_0$ has a negligible effect on $f_s(q,\Delta t)$ as long $1/q \ll l_0$, which is the regime typically probed in experiments \cite{cipelletti2003universal,bandyopadhyay2004evolution,robert2006glassy,chung2006microscopic}. 
By contrast, if the probed length scale $1/q$ is large enough to exceed the largest particle displacements, $f_s(q,\Delta t)$ is a Gaussian function of $q$ and takes the form
$f_s(q,\Delta t)\simeq \exp\left[-\frac{q^2\langle \Delta r^2 \rangle}{2d}\right]\simeq \exp\left[-\frac{q^2 l^{2-\alpha}_0 r_0(\Delta t)^{\alpha}}{2d}\right]$, where the first identity holds up to the second order in $q$ \cite{dhont1996introduction, giavazzi2018tracking}. The ISF is here still described by a compressed exponential function $f(q,\Delta t)\simeq e^{-[\Gamma(q)\Delta t]^\alpha}$;  the compressing exponent $\alpha$ is the same as that obtained for large $q$, the relaxation rate, however, follows a completely different dispersion relation $\Gamma(q)=(2d)^{-1/\alpha}(u_0/l_0)(l_0 q)^{\delta}$, with $\delta=2/\alpha$.

This model is fully consistent with our experimental findings ($\delta\simeq 1.6$ and $\alpha\simeq 1.2$) and provides the essential framework for the relation between dynamical characteristics in respectively reciprocal and direct space.

\subsection{\label{sec:tracking} Direct space: bubble dynamics}
\begin{figure*}
\centerline{\includegraphics[width=1.23\linewidth]{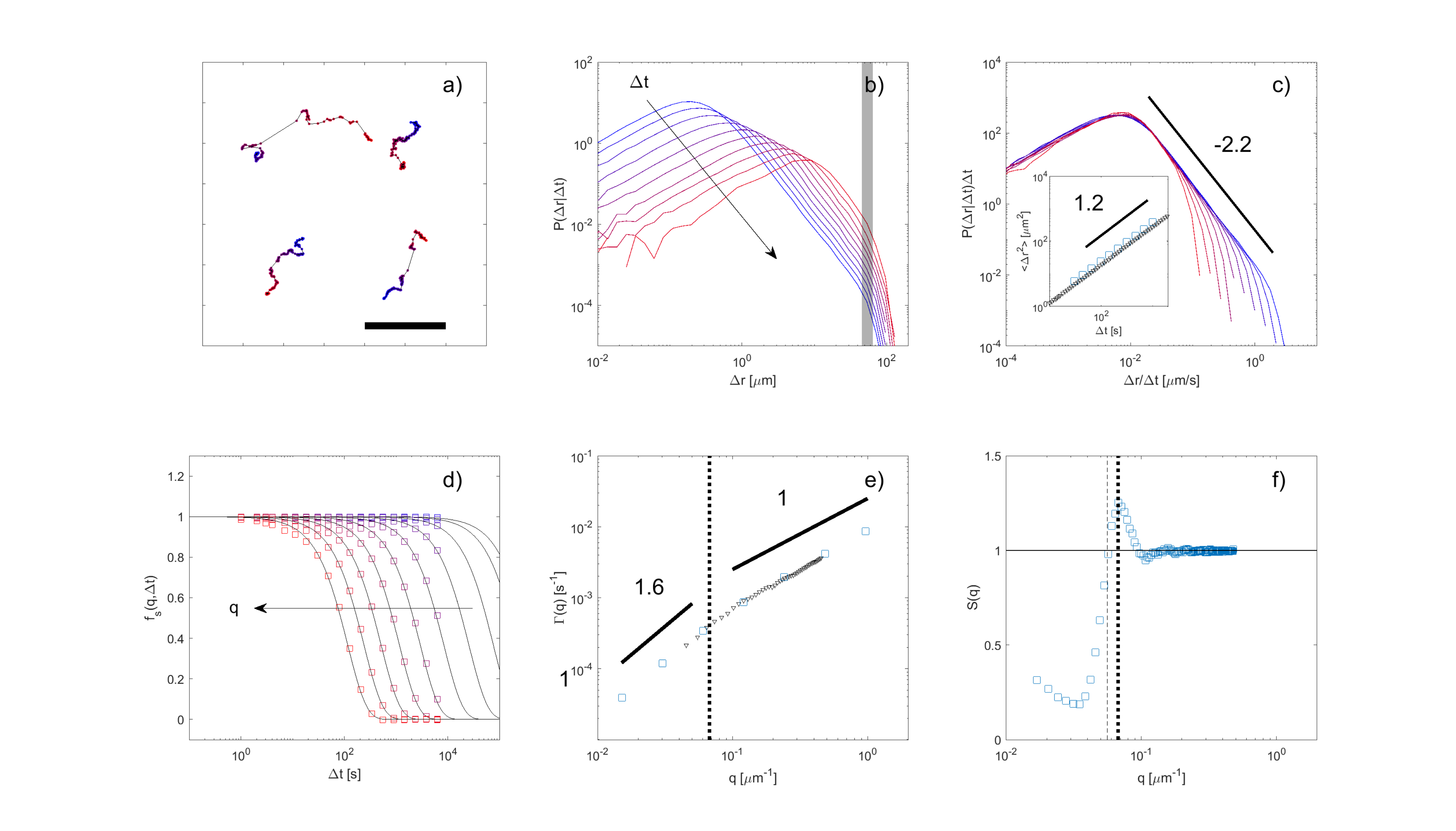}}
\caption{\label{fig3_1}
Real space analysis of the data obtained at $t_w=1.5 \cdot 10^4$ s.  a) Representative examples of bubble trajectories.  Each trajectory covers a time interval of 1400 s; consecutive dots are separated by a time interval of 10 s; the scale bar corresponds to 20 $\mu m$. 
b) Probability density  functions of bubble displacements $P(\Delta x,\Delta t)$ for $\Delta t$ ranging from 30s to 1000s. The grey area corresponds to a narrow interval around $60$ $\mu m$ corresponding to the cutoff length $l_0$. 
c) Scaled representation of the data shown in panel b); the horizontal and the vertical axis are rescaled with $\Delta t$ and $1/\Delta t$, respectively. Inset: Mean square displacement obtained from particle tracking are denoted as blue circles, those obtained from $f(q_1,\Delta t)$ are shown as black downward triangles.   
d) Self intermediate scattering functions obtained from particle tracking for $q$-values ranging from 0.15 $\mu m^{-1}$ to 78 $\mu m^{-1}$.  Continuous black lines are best fits to the data using compressed exponentials of form $\exp{\{-(\Gamma(q)\Delta t)^{\alpha}\}}$ with $\alpha=1.2$.
e) Comparison of the $q$-dependent relaxation rates obtained from PT (blue squares) and DDM (black downward triangles) analysis. The vertical dotted line denotes the position of the peak in $S(q)$.
f) Static structure factor $S(q)$ of the bubble centers. The vertical dashed line denotes the lowest $q$-value  at which $S(q_1)=1$ ($q_1=0.056$ $\mu m^{-1}$); the vertical dotted line denotes the peak position ($q=0.067$ $\mu m^{-1}$). 
}
\end{figure*}

To test this relation, we determine the bubble displacements in real space for a fixed foam age ($t_{w}=1.5 \cdot 10^4$s). A typical map of the trajectories obtained over a time interval of 1400s in steps of 10s is shown in Figure \ref{fig3_1}a. Each trajectory displays directional persistence, consistent with ballistic-like motion inferred from the linear dependence of $\Gamma(q)$ on $\Delta t$ observed at larger $q$. 
The PDF of particle displacements displays a well-defined peak for any given $\Delta t$, as shown in Figure \ref{fig3_1}b. At larger $\Delta r$ the PDF decreases as a power-law with an exponent $\alpha +1\sim 2.2$. For the smallest $\Delta t$ considered, this regime extends over about two decades, before being truncated at a cut-off length scale $l_0 \approx 60$ $\mu m$. The peak of the PDF systematically shifts to larger $\Delta r$ as $\Delta t$ is increased, while the cut-off length-scale is almost fixed at a value corresponding to approximately the characteristic bubble diameter $2R\simeq 62$ $\mu m$. 
The $\Delta t$-dependence of the PDF is fully consistent with ballistic-like motion. Indeed, a simple normalization of $\Delta r$ and of the amplitude of the PDF with $\Delta t$ leads to an excellent collapse of the data up to the cut-off length, as shown in Figure \ref{fig3_1}c. 
Moreover, the MSD (inset of Figure \ref{fig3_1}c, blue squares) scales as  
$\langle \Delta r^2 \rangle \sim \Delta t^{\alpha}$ with $\alpha$=1.2, in full agreement with the expectation from the truncated power law behavior of the PDF predicted by the simple model outlined in Subsection \ref{sec:model}. 

As a further consistency check, we evaluate the self intermediate scattering function $f_s(q,\Delta t)=\frac{1}{N}  \sum_{m=1}^N\langle{e^{-j\mathbf{q}\cdot(\mathbf[{r}_m(t_0+\Delta t) -\mathbf{r}_m(t_0)]}} \rangle_{|\mathbf{q}|=q,t_0}$.  Consistent with the results obtained in reciprocal space, $f_s(q,\Delta t)$ is well described by a compressed exponential with an exponent of $1.2$ (Figure \ref{fig3_1}d). Moreover, the  $q$-dependence of the relaxation rate $\Gamma_s(q)$ shown in Figure \ref{fig3_1}e clearly displays two distinct dynamic regimes, in excellent agreement with the analogous quantity obtained from DDM analysis of the same image sub-sequence. The quality of the agreement is actually somewhat surprising. It indicates that the decay of the ISF probed in DDM is dominated by its self-part; effects due to collective dynamics seem to be negligible over the whole $q$-range accessible in DDM.

We further exploit PT to calculate the static structure factor $S(q)$ of the bubble centers, which exhibits a well defined peak at $q\simeq 0.067$ $\mu m^{-1}$ $\simeq 2/R$ (Figure \ref{fig3_1}f). This $q$-value corresponds to the crossover wavevector separating the two dynamical regimes, which  supports the idea that the scale-dependent dynamics originates from a cut-off in the displacements that corresponds to the bubble length scale. Considering that the displacements of bubbles are determined by local stress imbalances that will occasionally exceed the yield conditions, this indicates that a new local stress configuration is only reached once the bubble has moved by its own diameter.




As a further test of the relation between reciprocal and direct space results we focus on the wavevector, denoted as vertical dashed line in Figure \ref{fig3_1} f), where $S(q_1)=1$. This $q$-value falls in the low $q$ dynamic regime, where $f(q,\Delta)$ is, in good approximation, a Gaussian function of $q^2$, suggesting that the single bubble MSD can be determined as $-4\ln[f(q_1,\Delta t)]/q_1^2$. As shown in the inset of Fig. \ref{fig3_1}c) this estimate is in very good agreement with the PT results. 

\section{Conclusions}
Our investigation on a coarsening foam reveals that bubble dynamics is governed by intermittent displacements that exhibit a persistent direction up to a given length scale. This cut-off length leads to distinct features in the dispersion relation of the relaxation rate $\Gamma (q)$ probed as a function of the wave-vector $q$ in reciprocal space. In our foam, the cut-off length corresponds to the bubble diameter 2$R$; at $1/q$–values smaller than 2$R$, $\Gamma (q)$ scales linearly with $q$ consistent with the results obtained for aging colloidal gels \cite{cipelletti2000universal}.  By contrast, for  $1/q > 2R$, we find a scaling of $\Gamma (q) \sim q^{2/\alpha}$, with $\alpha$ = 1.2. We show that introducing a cut-off length into the models proposed in \cite{bouchaud2001anomalous,duri2006length,bouchaud2008anomalous} naturally accounts for this behaviour; in addition, it explains the compressed exponential relaxation of the intermediate scattering function  
$f(q,\Delta t) = \exp{[-(\Gamma(q)\Delta t)^{\alpha}]}$
observed in both $q$-regimes.

To put our results into general context, let us note that the magnitude of the compressing exponent $\alpha$ observed in our experiment is significantly smaller than  $\alpha\simeq1.5$ reported in dynamic light scattering studies on other systems  \cite{cipelletti2003universal}.
According to the mean-field arguments presented in ref. \cite{cipelletti2003universal} 
the exponent $\alpha$ is determined by the ratio $d/\beta$, where $\beta$ is the exponent of the leading term in the decay of the displacement field $u(r)$ generated by a single dipolar plastic event occurring at the origin $u(r)\sim u^{-\beta}$. In three dimensions ($d=3$) $\beta=2$, leading to $\alpha=3/2$ \cite{cipelletti2003universal}. 
The deviation of the observed exponent ($\alpha \simeq 1.2$) from this value could be due to the geometry of our sample. The thickness of the sample is indeed much larger than the average bubble radius (at least 150 times), the observation plane, however, coincides with one of the confining walls. At this 3D semi-infinite condition, the far-field decay of the displacement field generated by a single plastic event should be that of the 3D unbounded case, but due to events occurring close to the wall we expect significant near-field corrections \cite{picard2004elastic}. These could lead to an effective, faster-then-quadratic decay of the displacement field, which would explain the deviation of the observed exponent from the mean-field value expected in the 3D unbounded case.

With respect to the origin of dynamics in foams, our experiments unambiguously show that the constantly renewed mechanical constraints imposed by the coarsening process are the cause for persistent dynamics.  This is evidenced by a direct correlation between the age dependent relaxation rates and the strain rates imposed by the increase in bubble size.  On a microscopic scale, the persistence in direction of bubble displacements from one intermittent event to another is consistent with previous observations, reporting that subsequent bubble rearrangement events preferentially occur at the same location \cite{sessoms2010unexpected}.  Considering that an event is triggered by local stress-imbalances, these findings indicate that an event does not necessarily rejuvenate the stress configuration.  Indeed, we can argue that the bubbles start to move when the net local stress exceeds the yield stress, and that they will stop moving once the local stress is below the yield stress again.   This entails that the event location remains among the most fragile regions of the system, and that the direction of the net local stress is not significantly changed after an event, consistent with the observed behaviour.

Our results significantly contribute to the understanding of dynamics that is driven by internal stresses. They provide clear evidence of the driving mechanism for intermittent bubble rearrangements in foam and they unveil a limit for directionally-persistent displacements. We believe that investigations specifically aiming to explore the source for stress-driven dynamics and the existence of a cut-off length for directionally-persistent displacements in other systems would be highly beneficial to fully establish the mechanisms of stress-driven dynamics.           

In this context, investigations on cell tissues appear promising. Indeed, foams and cell tissues exhibit  similar tessellation patterns \cite{thompson1942growth,gibson1999cellular,siber2017cellular}, and it has been shown that simple models, originally developed to describe the configuration of foams and other jammed systems, can be extended to rationalize experimental results on cell tissues \cite{farhadifar2007influence,bi2014energy,bi2015density,giavazzi2018flocking}  \cite{park2015unjamming,malinverno2017endocytic}.
More importantly, both tissues and foams are non-equilibrium, slowly evolving systems, displaying heterogeneous, intermittent, super-diffusive dynamics and long-range correlations \cite{duri2009resolving,sessoms2010unexpected,angelini2011glass,giavazzi2018tracking}.
The source for persistent dynamics are different: in foams, the source is the coarsening process, which is induced by the pressure difference between differently sized gas bubbles, in cell tissues, energy is continuously injected at the single cell level, cell motility and proliferation being major drivers of structural reorganisation \cite{trepat2018mesoscale,berthier2009glassy}.
Only recently, a study similar that presented here for foams, using a combination of real- and reciprocal-space diagnostic tools typically used in soft condensed matter physics have been applied to cell tissues  \cite{angelini2011glass,zehnder2015multicellular,giavazzi2017giant,giavazzi2018tracking}, unveiling their potential in providing a robust multi-scale description, which should facilitate the theoretical description of these systems.
This strategy appears particularly promising to establish the link between spatial structure and dynamics, which represents one of the major challenges in understanding the behaviour of complex active systems close to dynamical arrest \cite{Janssen2019}.

\ack{}
We thank Luca Cipelletti for stimulating discussions. We acknowledge funding from the Associazione Italiana per Ricerca sul Cancro (MFAG 2018-22083) (FG), the Swiss National Science Foundation grant-number 200021-172514 (VT) and the European Space Agency (MAP Contract 4000128933 - TechNes) (RC).

\section*{References}
\bibliography{biblio_foam}
\bibliographystyle{iopart-num}

\end{document}